\documentclass{llncs}

\usepackage[margin=1in]{geometry}
\usepackage{multirow}
\usepackage{url}
\usepackage{graphicx}

\title{Enforcing Privacy in Cloud Databases}

\author{Somayeh Sobati Moghadam \and J\'{e}r\^{o}me Darmont \and G\'{e}rald Gavin}

\institute{Universit\'{e} de Lyon, Lyon 2, Lyon 1, ERIC EA3083\\
	5 avenue Pierre Mend\`{e}s France -- 69676 Bron Cedex -- France\\
	\email{ssobati@eric.univ-lyon2.fr, jerome.darmont@univ-lyon2.fr, gerald.gavin@univ-lyon1.fr}
}

\begin{document}

\maketitle

\begin{abstract} \hyphenation{now-ad-ays}
Outsourcing databases, i.e., resorting to Database-as-a-Service (DBaaS), is nowadays a popular choice due to the elasticity, availability, scalability and pay-as-you-go features of cloud computing. However, most data are sensitive to some extent, and data privacy remains one of the top concerns to DBaaS users, for obvious legal and competitive reasons.
In this paper, we survey the mechanisms that aim at making databases secure in a cloud environment, and discuss current pitfalls and related research challenges.	
\end{abstract}

\section{Introduction}

\hyphenation{par-ad-igm}
\hyphenation{data-bases}
	
Cloud computing offers a variety of services via a pay-per-use model on the Internet. The flexibility that cloud computing offers is very appealing for many organizations, especially mid-sized and small ones, because it provides reduced start-up costs and means to financially cope with variations in system usage. Outsourcing data to the cloud is particularly interesting \cite{xiong2007preserving}.
However, some data are especially sensitive, e.g., personal data, health-related data, business data, and generally data used in decision-support processes. Outsourcing a database in the cloud raises security issues, some related to cloud architectures (e.g., untrusted service providers, curious cloud employees...), and others related to such concerns as data privacy, integrity and availability. With increasingly sophisticated internal and external cloud attacks, traditional security mechanisms are no longer sufficient to protect cloud databases \cite{yuhanna2009your}. 	

\begin{figure}[hbt]
	\centering
	\includegraphics[width=9cm]{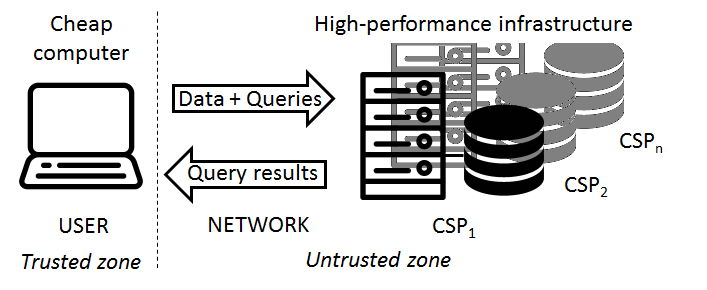} 
	\caption{Database outsourcing scenario}
	\label{fig:scenario}
\end{figure}

Let us consider a Database-as-a-Service (DBaaS) scenario (Figure~\ref{fig:scenario}) where a user outsources a database at one or more Cloud Service Providers' (CSPs). The objective is to eliminate storage and minimize computation at the user's to take full advantage of cloud benefits. Yet, anything beyond the user is considered untrusted. CSPs might indeed be honest but curious, i.e., read the user's data, or even be malicious or maliciously hacked, i.e., alter data or provide fake query results. Network transactions are also considered unsafe. The user must thus protect sensitive data and queries before sending them to CSPs, and safely reconstruct query results ``at home''.

In this paper, we survey the security mechanisms that may be exploited in our cloud database scenario, particularly in terms of privacy. Another recent survey intersects ours \cite{sigmodsurvey16}, but more deeply focuses on cryptography, while we adopt a broader scope on security, put more emphasis on querying efficiency and survey non-cryptographic methods (Section~\ref{Section.NonCryptMethods}). We also mostly target database practitioners and researchers with no background in cryptography. Moreover, we review cryptographic methods that are not covered in \cite{sigmodsurvey16}, i.e., secret sharing, Private Information Retrieval (PIR) and oblivious RAM (ORAM) schemes. 
We classify cryptographic tools that can be exploited within cloud database scenarios into secret sharing schemes (Section~\ref{Section.secretsharing}), index-based methods (Section~\ref{Section.IndexBasedMethods}) and secure databases (Section~\ref{Section.SecureDBs}). 
Finally, we conclude this paper by a global discussion (Section~\ref{section.GlobalDiscussion}).


\section{Non-Cryptographic Methods}
\label{Section.NonCryptMethods} 

\subsection{Differential Privacy}

Differential privacy aims at protecting data privacy when performing statistical queries \cite{DBLP:conf/icalp/Dwork06}. While global statistics are public,
individual data must remain private. To achieve this goal, a noise term is added to statistical query results, e.g., to an average salary, thus preventing the computation of individual salaries. It would indeed be easy to compute a new salary that has just been added in the database, knowing averages $avg$ and the number of records $n$ in the dataset: $avg_{n+1} \times (n+1) -  avg_{n} \times (n)$.

A randomized algorithm $A$ enforces $\epsilon$-differential privacy if and only if, for any
two databases $DB_1$ and $DB_2$ that differ on exactly one record, the ratio between the probability that $A$ outputs $O$ on $DB_1$ and $DB_2$ is bounded
by a constant: $P(A(DB_1)=O) \div P(A(DB_2)=O) \leq e^\epsilon$ \cite{DBLP:conf/sigmod/YangZMWX12}. The tradeoff is that the smaller $\epsilon$ is, the better privacy is enforced, but the worse accuracy is. Thus, one major challenge is reducing the amount of noise while still satisfying differential privacy.
Yet, in a DBaaS context, a curious CSP would still have access to fine-grained data, which is incompatible with our data outsourcing scenario, where we aim at protecting data event from the CSP.

\subsection{Data Anonymization}
\label{Section.NonCryptMethods.Data Anonymization} 

\hyphenation{an-on-ym-iz-at-ion}
\hyphenation{data-base}
\hyphenation{an-on-ym-iz-ed}
\hyphenation{quer-ies}

Data anonymization 
 irreversibly modifies data in a way that prevents the identification of sensitive information, while allowing querying data for releasing useful statistical information \cite{DBLP:reference/crypt/Dwork11}. 	
Some database management systems (DBMSs) natively include anonymization schemes. For example, Oracle Data Masking Pack  provides data masking for various types of data, replacing real data with realistic-looking values \cite{oracle}. 
Obscuring query processing and results may also be achieved either by rejecting queries leading to privacy disclosure or through methods such as k-anonymity and its variants t-closeness and l-diversity, 
which transform $k$ distinguishable records  into $k$ indistinguishable records \cite{DBLP:journals/ijufks/Sweene02}.

In a DBaaS context, privacy-preserving queries in a distributed environment can be achieved by table perturbation and reconstruction \cite{DBLP:conf/sigmod/AgrawalST05}. Perturbation randomly replaces an element in a table by another with probability $p$. Then, reconstruction can estimate \texttt{COUNT} queries over the perturbed table.  Unfortunately, other aggregation functions are not supported. Similarly, random data distortion techniques may be used, e.g., the zero-sum method    provides an accurate estimation of summation for range queries \cite{DBLP:journals/kais/SungLXN06}, but it induces a trade-off between privacy and accuracy. Finally, 
adding different amounts of noise to query answers can be used, e.g., 
with iReduct \cite{DBLP:conf/sigmod/XiaoBHG11}. iReduct initially estimates query answers and iteratively refines its estimates to minimize relative errors.

Data anonymization is less complex than encryption and straightforwardly allows querying anonymized data. However, it 
typically reduces data granularity, which may cause a loss of effectiveness and correctness in computation \cite{DBLP:series/ads/AggarwalY08a}. 
Moreover, the various anonymization methods have different impacts on  data utility. 
As a result, the same anonymized table may provide accurate answers to some queries and inaccurate results to others. 
More importantly, data anonymization cannot provide an adequate level of security since private personal data can be re-identified by an adversary who has some knowledge about data 
\cite{montjoye13}.

\subsection{Data Fragmentation}
\label{Section.NonCryptMethods.DataFrag}

In data fragmentation \cite{DBLP:conf/cidr/AggarwalBGGKMSTX05}, 
data are assumed not to be sensitive \emph{per se}. What is sensitive is their association with other data. Privacy is thus guaranteed by concealing such associations with respect to a predefined set of security constraints  that express restrictions on one or more attributes in a table \cite{DBLP:journals/jcs/CirianiVFJPS11,DBLP:conf/compsac/HadaviNJD12}. 
For instance, given a $Patient$ table, constraint $C=\{Name, Illness\}$ 
indicates that associations between patient names and illnesses should not be disclosed. 
Then, the table is split into fragments such that attributes listed in a constraint belong to different fragments. For instance, table $Patient$ would be split in two fragments $Patient_1$ and $Patient_2$, with $Name \in Patient_1$ and $Illness \in Patient_2$. Yet, most data fragmentation approaches apply to numerical data and specific methods most be used to handle categorical data \cite{DBLP:conf/mdai/RicciD016}.

In a cloud context, data can be partitioned at independent CSPs' with respect to security constraints. When a query is issued, an appropriate subquery is transmitted to each CSP, then the results are pieced together at the user's. Moreover, intrinsically sensitive attributes such as social security numbers are stored locally at the user's.
Eventually, fragmentation-based approaches yield little overhead on query computation, but are vulnerable in cases of CSP collusion and CSP inference on data updates \cite{DBLP:conf/esorics/HadaviDJCG12}. Additionally, retaining sensitive data at the user's requires local storage capacities, which is incompatible with our data outsourcing scenario.

\section{Secret Sharing-Based Methods}
\label{Section.secretsharing}

Secret sharing is a particular cryptographic method introduced by Shamir in which  
a secret piece of data is mathematically divided into so-called shares that are stored at $n$ participants' \cite{shamir1979share}. One single participant has no means to reconstruct the secret. A subset of $k \leq n$ participants is indeed required to reconstruct the secret, providing perfect theoretical privacy when at most $k-1$ participants collude, i.e., exchange shares. Moreover, computations can run directly on shares, outputting unintelligible results that can only be put together through $k$ participants. Finally, up to $n - k$ participants may disappear without compromising data availability; and message authentication code or signature can be applied to guarantee data integrity. Thus, secret sharing is a promising solution to security challenges in cloud data outsourcing \cite{DBLP:conf/esorics/HadaviDJCG12} since one can easily imagine participants being CSPs. 
In the following subsections, we review the secret sharing-based approaches that aim at outsourcing databases, and then discuss them globally.

\subsection{Verifiable Secret Sharing}
\label{sec:VerifiableSecretSharing}


Thompson et al.'s scheme allows participants to collaboratively compute aggregation queries without gaining knowledge of intermediate results \cite{thompson2009privacy}. A lightweight cryptographic scheme is introduced for privacy-preserving computation
and verification of \texttt{SUM} and \texttt{AVG} aggregation queries.  Moreover, users can verify query results with the help of signatures, while the data values contributing to the results are kept secret from both users and the CSPs. The query issuer indeed interacts with a single CSP
to obtain aggregation results and can verify whether  the CSP
returns correct results.
However, aggregation queries other than \texttt{SUM} and \texttt{AVG} cannot be computed with this scheme. 		


Attasena et al. specifically target cloud data warehouses through a flexible, verifiable secret sharing scheme named fVSS \cite{DBLP:conf/dolap/AttasenaHD14}. fVSS minimizes shared data volume, provides inner and outer data verification to check data correctness and the honesty of CSPs, improves the ability to update shares in case of CSP failure, and adjusts share volume with respect to CSP pricing policies. Moreover, in addition to queries explicitly handled by previous schemes, fVSS also allow grouping queries that are ubiquitous in On-Line Analysis Processing (OLAP). However, although some queries can be computed directly over shares (e.g., exact match queries), others require that some or all data are decrypted first (e.g., range queries). 


Wang et al.  propose a framework for secure and efficient query processing of
relational data in the cloud that allows exact match and range queries, as well as updates \cite{wang2011comprehensive}. B+-tree indexes are also built to optimize query response. 
Both data and indexes are organized
into matrices, encrypted and stored at CSPs'. Additionally, data integrity is achieved by using checksum and
an index structure. This framework is robust against statistical attacks.

Statistical attacks refer to an adversary obtaining some information about ciphertexts, i.e., encrypted data, through prior knowledge about plaintexts, i.e., clear data \cite{DBLP:conf/esorics/HadaviDJCG12}.
For instance, the adversary may know plaintext distribution or frequency. Then, by extracting statistics from ciphertexts, the adversary can infer ranges containing dense data or highlight ciphertexts bearing the same frequency as plaintexts.

\subsection{Order-preserving Secret Sharing}
\label{sec:OrderPreservingSecretSharing}


Agrawal et al. propose a complete approach to execute exact match, range, and aggregation queries over shares outsourced at multiple CSPs \cite{DBLP:conf/icde/AgrawalAEM09}. 
	Original data are divided using order-preserving polynomials such that the order of shares is the same as that of original data. However, while this solution allows efficiently processing any kind of queries, including updates, it is susceptible to statistical attacks \cite{DBLP:conf/esorics/HadaviDJCG12}.


Hadavi et al. introduce a framework
to provide data privacy based on threshold
secret sharing \cite{hadavi2010secure}.  
First, secret values are encrypted by an Order Preserving Encryption (OPE) scheme (Section~\ref{Section.SecureDBs.CryptDB}). Then, a B+-tree is built over ciphertexts and sent to an index server. The user receives query responses, including record numbers, from the index server and can then request these records from the CSPs.
As Agrawal et al.'s scheme, this approach supports different kind of queries over shares, including exact match, range and aggregation queries, as well as updates. Moreover, it provides stronger security than Agrawal et al.'s scheme. It is indeed secure against frequency attacks, and an extension uses distribution perturbation to improve its robustness against statistical attacks in general \cite{DBLP:conf/esorics/HadaviDJCG12}; but at the price of a higher computational overhead.

\subsection{Discussion}
\label{Section.secret sharing.Disc}	

Table~\ref{my-label-2} provides a comparison of secret sharing-based database outsourcing methods with respect to the queries that can run directly on shares and security features beyond privacy and availability, i.e., integrity checks and robustness against statistical attacks.   	

\begin{table*}[hbt]
	\centering
	\caption{Comparison of secret sharing-based methods}
	\label{my-label-2}	
	\begin{tabular}{|l|c|c|c|c|c|c|c|}\hline
		& \multicolumn{4}{c|}{Allowed queries}  & \multicolumn{2}{c|}{Additional security features}                                                 \\ \cline{2-7}
		& Exact match & Range & Aggregate  & Update & Integrity & Statistical attacks \\
		\hline
		Thompson et al. \cite{thompson2009privacy}    & No                   & No               & SUM/AVG     & No      & Signature          & Not robust                               \\
		Attasena et al. \cite{DBLP:conf/dolap/AttasenaHD14}    & Yes                  & Yes              & Yes         & Yes     & Signature          & Not robust   \\		
		Wang et al. \cite{wang2011comprehensive}       & Yes                  & Yes              & Yes         & Yes     & Checksum           & Robust                             \\ \hline
		Agrawal et al. \cite{DBLP:conf/icde/AgrawalAEM09}    & Yes                  & Yes              & Yes         & Yes     & None               & Not robust                              \\
		Hadavi et al. \cite{hadavi2010secure}     & Yes                  & Yes              & Yes         & Yes     & None               & Robust                             \\
		\hline                          
	\end{tabular}
\end{table*}

Despite secret sharing's benefits, it is not trivial to process some queries directly over shares, especially queries requiring data ordering. Obviously, it is not efficient to send back all shares in response to a query, and execute the query over reconstructed values at the user's.  
However, the techniques that allow directly processing such queries as range queries over shares reveal some information about plaintexts, e.g., duplicates \cite{DBLP:conf/icde/AgrawalAEM09}.  
Moreover, since every secret is shared $n$ times, global share size can be quite large. 
Communication between the user and multiple CSPs is not optimal in terms of bandwidth resources either. 
As a result, storage and communication overhead of secret sharing-based approaches is remarkably high for moderately large databases \cite{DBLP:conf/crypto/Krawczyk93}.

\section{Index-Based Methods}
\label{Section.IndexBasedMethods}

In databases, data encryption is usually managed at the tuple level \cite{DBLP:journals/tods/DavidaWK81}, which prevents any computation over ciphertexts. Thus, indexes based on plaintexts are stored together with the encrypted database to 
help return ciphertexts in response to queries. 
We distinguish three types of index-based methods, namely bucketization, order preserving indexing and indexes used in Searchable Encryption (SE) schemes. We review them in the following subsections before providing a global discussion.

\subsection{Bucketization-Based Indexing}
\label{Section.IndexBasedMethods.Bucket}

\hyphenation{da-ta}

Bucketization-based indexing involves dividing data into buckets and providing explicit labels for each bucket \cite{DBLP:conf/ccs/MykletunT06}. 
The domain of an attribute is partitioned into a set of non-overlapping buckets. 
Labels may preserve the order of values in the original domain or not. They are stored along with encrypted tuples. 
Such indexing allows exact match, range (if data order is preserved) and join queries, but also induce false positives in query answers. 
Thus, query post-processing is needed at the user's to filter out false positives \cite{DBLP:conf/ccs/SamaratiV10}.


Hacig{\"{u}}m{\"{u}}s et al. partition data as in histogram construction, e.g., by equi-depth and equi-width partitioning \cite{DBLP:conf/sigmod/HacigumusILM02}. Then, it assigns a random tag to each bucket. 
Any table 
$T(A_1, A_2,...,A_n)$ from a database is stored at the CSP's as $T^S(etuple,$ $A_1^S, A_2^S, ..., A_n^S)$, where $etuple$ is the encrypted tuple and each $A_i^S$ is the index of attribute $A_i$. 
Each query is rewritten into server-side and user-side subqueries $Q^S$ and $Q^C$, respectively. $Q^S$ is executed by the CSP over ciphertexts using indexes $A_i^S$. The result of $Q^S$ is then sent back to the user, who decrypts it and executes $Q^C$ to filter out false positives. 
Query rewriting requires maintaining metadata, including bucket labels. 

With the help of an homomorphic function, this approach is extended to support aggregation queries over ciphertexts \cite{DBLP:conf/dasfaa/HacigumusIM04}. 
The homomorphic encryption function is based on the Privacy Homomorphism (PH) scheme \cite{rivest1978data}, which relies on the difficulty of factoring large composite integers, just like the famous Rivest-Shamir-Adleman (RSA) public-key cryptosystem.  	
Unfortunately, Mykletun and Tsudik  demonstrate  
that the CSP can obtain plaintexts with access to ciphertexts only \cite{DBLP:conf/ccs/MykletunT06}.


Based on their rebuttal of Hacig{\"{u}}m{\"{u}}s et al., 
Mykletun and Tsudik  propose a simple alternative for supporting aggregation queries \cite{DBLP:conf/ccs/MykletunT06}. The user precomputes aggregation values (e.g., \texttt{SUM} and \texttt{COUNT}) for each bucket, encrypts and stores them at the CSP's. 
Moreover, instead of using the PH scheme, Mykletun and Tsudik use provably secure additive homomorphic encryption schemes such as Paillier's \cite{paillier1999public} and El Gamal's \cite{DBLP:journals/tit/Elgamal85}.
Precomputing aggregations decreases security risks, but requires extra storage and makes updates more complex. Updates must indeed be executed in two steps: 1) actual data update and 2) update of related precomputed aggregates in a bucket.


Hore et al. also address shortcomings of Hacig{\"{u}}m{\"{u}}s et al.'s method. 
They notably optimize the accuracy of range queries to minimize false positives in query results \cite{DBLP:conf/vldb/HoreMT04}. They also introduce a re-bucketization technique, in which the user can fine-tune bucketization to achieve a desired level of privacy. 
Moreover, they propose a new method for answering range queries on multidimensional data \cite{DBLP:journals/vldb/HoreMCK12}. Range queries over multiple attributes, e.g., $age < 20$ and $salary > 25$k, are allowed, while  
minimizing the cost of multidimensional bucketization. Yet,
a threshold is defined to help the user control the trade-off between risk of data disclosure and cost.

\subsection{Order-Preserving Indexing}
\label{Section.IndexBasedMethods.OP}	

\subsubsection{Order Preserving Encryption Scheme (OPES)}
\label{Section.IndexBasedMethods.OP.Agrawal}

Agrawal et al.'s OPES is an OPE indexing scheme that supports range and equality queries over integers \cite{DBLP:conf/sigmod/AgrawalKSX04}.
OPES transforms plaintexts with an order preserving function so that transformed values (e.g., index values) follow a target distribution. Comparison operations can be directly applied at the CSP's without inducing spurious tuples nor false positives.  
However, this scheme has been demonstrated to be vulnerable to statistical attacks \cite{liu2014new}.

\subsubsection{OPE with Splitting and Scaling (OPESS)}
\label{Section.IndexBasedMethods.OP.Wang}

The OPESS scheme encrypts XML databases \cite{DBLP:conf/vldb/WangL06}. Wang et al. adopt splitting and scaling techniques to create index values following a uniform distribution. Plaintext order is preserved over indexes. Moreover, identical clear values are transformed into different indexes so that this scheme is robust against statistical attacks. However, this scheme flattens the frequency distribution of index values.

\subsubsection{B+tree indexing}
\label{Section.IndexBasedMethods.OP.Elovici} 

Shmueli et al. \cite{DBLP:conf/dbsec/ShmueliWEG05} 
and Damiani et al. \cite{DBLP:conf/ccs/DamianiVJPS03} use B+-trees built on database plaintext attribute values to preserve order in secure environments. 
B+-trees must either be stored in a trusted machine \cite{hadavi2010secure} or be encrypted at the CSP's, where each B+-tree is stored in a table with two attributes: node identifier and node content.
In addition to ordering, B+-tree indexes support exact match, range and grouping queries, as well as predicates such as \texttt{LIKE}. For example, to execute a range query, the user sends a sequence of queries until reaching the leaf corresponding to the range's lower bound. Then, the node identifier helps retrieve all the tuples belonging to the range. The advantage of such indexing is that index content is not visible to the CSP and reveals no information about underlying plaintexts \cite{DBLP:conf/ccs/DamianiVJPS03}.

\subsection{Searchable Encryption}


SE allows the CSP to run keyword-based searches on encrypted data \cite{DBLP:conf/sp/SongWP00} that are particularly suitable to data outsourcing \cite{DBLP:journals/csur/BoschHJP14}. 
Considering a set of documents $\{D_i\}_{i=1,n}$ and an index of keywords $\{w_j\}_{j=1,m}$ describing the documents, users encrypt both documents $D_i$ with any secure encryption scheme using a key $\mathcal{K}_{Enc}$ and keywords $w_j$ with a searchable scheme using a key $\mathcal{K}_{Index}$.
The encrypted documents and index are then outsourced. 
When searching for documents containing some keywords, the user sends a so-called trapdoor encapsulating the keywords to the CSP \cite{DBLP:books/sp/securecloud14/SunLHL14}. Then, the CSP can search the encrypted index 
and the trapdoor to find the corresponding documents and send them back to the user. 
Both symmetric (private) and asymmetric (public) key encryption 
can be used to build symmetric SE (SSE) and asymetric SE (ASE) schemes, respectively \cite{DBLP:books/sp/securecloud14/SunLHL14}. 
ASE schemes support various query types such as range and subset queries, but are computationally intensive. SSE is more efficient than ASE, but supports fewer query types. 
SE induces a trade-off between security, efficiency and query expressiveness. SE schemes with higher levels of security induce higher complexity, while SE schemes supporting more query types are either less secure and/or less efficient \cite{DBLP:journals/csur/BoschHJP14}.  
Moreover, most SE schemes reveal access patterns, i.e., which documents contain a keyword. Only
techniques based on PIR or ORAM do not.

PIR \cite{DBLP:conf/eurocrypt/CachinMS99,DBLP:conf/acisp/Chang04,DBLP:conf/fc/LueksG15} 
enables a user to retrieve data from an outsourced database while preventing
	the CSP from learning any information about retrieved data \cite{DBLP:journals/jacm/ChorKGS98}, i.e., PIR enforces query  privacy.
Unfortunately, in a single-server setting, the only thing a user can do is retrieving the whole database, which induces communication overhead and annihilates the benefits of outsourcing. However, in a multiple-server setting where copies of the database are stored at $k$ non-communicating/colluding CSPs, a user
	can hide queries by querying each server for a part of data, so that no server knows the
	whole query. 

ORAM allows reading and
	writing to memory without revealing access patterns to the CSP \cite{DBLP:journals/jacm/GoldreichO96}. In ORAM schemes, a user stores encrypted data at the CSP's and continuously shuffles and re-encrypts	data as they are accessed \cite{DBLP:conf/ccs/StefanovDSFRYD13}. 
	Let $P=(q_1,\dots,q_n)$ be an access pattern. The shuffling process induces the transformation of each query $q_i$ into multiple queries, producing a new access pattern $P^\prime$. An ORAM protocol is secure if two access patterns ORAM($P$) and ORAM($P^\prime$) are computationally indistinguishable.
	ORAM can be implemented using symmetric or fully homomorphic encryption (Section~\ref{Section.SecureDBs.CryptDB}). An alternative solution for hiding access patterns is to frequently send fake queries to CSPs
	to prevent any adversary from inferring correlations between frequently queried data  \cite{DBLP:conf/sigmod/MavroforakisCOK15}. Yet, generating fake but realistic-looking queries is a challenge.
	
Unfortunately, a common limitation of PIR and ORAM schemes is a prohibitive query execution time \cite{DBLP:journals/tissec/WilliamsS13}.	

\subsection{Discussion}
\label{Section.IndexBasedMethods.Disc}

Table \ref{my-label} provides a comparison of index-based methods with respect to the query types they allow and whether they require a post-processing step to eliminate false positives.

\begin{table*}[hbt]
	\centering
	\caption{Comparison of index-based methods}
	\label{my-label}
	\begin{tabular}{|l|c|c|c|c|}
		\hline
		& \multicolumn{3}{c|}{Allowed queries}           &  \\     \cline{2-4}                                                  
		& Exact match & Range & Aggregation & Post-processing\\
		\hline
		
		Hacig{\"{u}}m{\"{u}}s et al. \cite{DBLP:conf/sigmod/HacigumusILM02}                                           & Yes       & Yes           & Yes         & Yes                                                                                                                                   \\
		Mykletun \& Tsudik \cite{DBLP:conf/ccs/MykletunT06}                                         & No       & No           & Yes         & No                                                                \\
		Hore et al. \cite{DBLP:conf/vldb/HoreMT04}                                      & Yes      & Yes          & No          & Yes                                                                                                                                   \\ \hline
		Agrawal et al. \cite{DBLP:conf/sigmod/AgrawalKSX04}                                     & Yes      & Yes          & No          & No                                                                                                                                      \\
		Wang et al. \cite{DBLP:conf/vldb/WangL06}                                            & Yes      & Yes          & No          & No                                                                                                                                  \\ 
		Shmueli et al. \cite{DBLP:conf/dbsec/ShmueliWEG05}, 
Damiani et al. \cite{DBLP:conf/ccs/DamianiVJPS03} & Yes      & Yes          & No          & No                                                                                                                                \\ \hline
		Searchable Encryption \cite{DBLP:conf/eurocrypt/CachinMS99,DBLP:conf/acisp/Chang04,DBLP:journals/jacm/GoldreichO96,DBLP:conf/fc/LueksG15,DBLP:conf/sp/SongWP00,DBLP:books/sp/securecloud14/SunLHL14} & Yes & No & No & No \\
		\hline              
	\end{tabular}
\end{table*}

When defining an indexing method, it is important to consider two conflicting requirements. 
On one hand, the index should be related to the data well enough to allow efficient query execution. On the other hand, this relationship between plaintexts and the index should minimize the risk of any disclosure or loss of privacy \cite{DBLP:conf/sdmw/EloviciWSG04,DBLP:conf/ccs/SamaratiV10}. For example, in bucketization-based indexing, decreasing the number of buckets impairs performance, while a larger number of buckets increases the risk of data disclosure. 
In our database outsourcing scenario, a critical drawback of bucketization-based indexing is the loss of data granularity, which prevents grouping operations. The CSP can indeed not distinguish between tuples in buckets and the user has to filter intermediate results sent by the CSP to reconstruct the global result. Hence, bucketization-based methods induce computational overhead at the user's, too. Such an overhead can be high, especially for queries that return a large number of encrypted tuples \cite{DBLP:journals/pvldb/TuKMZ13}, e.g., grouping queries running on fine-grained data.
			
Finally, although index-based approaches are quite popular in cloud data outsourcing due to their efficiency \cite{DBLP:conf/esorics/HadaviDJCG12,DBLP:conf/ccs/SamaratiV10}, their main limitation lies in data update. Typically, such methods but SE exploit the distribution of plaintexts, while update operations may change it, making index regeneration unavoidable \cite{DBLP:conf/esorics/HadaviDJCG12}. As a result, index-based solutions but SE are suitable for read-only data. Yet, SE schemes are either too costly or too limited in query expressiveness to be used in practice.

\section{Secure Databases}
\label{Section.SecureDBs}


\subsection{CryptDB}
\label{Section.SecureDBs.CryptDB}

\hyphenation{sche-me}

CryptDB  is a pioneer system that allows efficient SQL query processing over ciphertexts into a DBMS \cite{DBLP:conf/sosp/PopaRZB11}. 
The properties of a cryptographic scheme determine the kinds of queries that can be directly executed over ciphertexts. Thus, CryptDB implements several schemes with respect to different user-determined security requirements and query needs. 
Thus, we first describe these cryptographic schemes, and then we detail CryptDB's architecture.

\subsubsection{Query-Aware Encryption Schemes}
\label{Section.SecureDBs.CryptDB.QAEncryption}

\paragraph{Random Encryption (RND)}
\label{RND}

RND schemes are the strongest security schemes. They indeed guarantee semantic security, i.e., it is computationally impossible to distinguish two ciphertexts. For instance, let $x$ be a plaintext value and $E$ an RND encrypting function. If, using the same encryption key, $e_1= E(x)$  and $e_2=E(x)$, then  with high probability, $e_1\neq e_2$. 	  
However, RND schemes do not allow any computations nor queries over ciphertexts. They are only designed for safe storage.

\paragraph{Homomorphic Encryption (HE)}
\label{HE}

\hyphenation{proh-ib-it-iv-ely}

HE allows performing arbitrary arithmetic operations over ciphertexts without decryption \cite{gentry2009fully} while still providing semantic security.  		
For instance, with an additive HE scheme, for any two encryptions $E(x)$ and $E(y)$, there exists a function $f$ such that $f(E(x), E(y))=E(x+y)$.	
Fully homomorphic encryption (FHE) is prohibitively slow and requires so much computing power that it cannot be used in practice as of today.
However, partially homomorphic encryption (PHE)  is efficient for specific operations and can be used in practice. PHE allows either addition or multiplication over ciphertexts and guarantees semantic security. Paillier's  \cite{paillier1999public} and El Gamal's \cite{DBLP:journals/tit/Elgamal85} are examples of PHE schemes. For instance, with Paillier's PHE, 
the product of two encryptions encrypts the sum of the encrypted values, 
i.e., $E(x) \times E(y) = E(x+y)$.

\paragraph{Deterministic Encryption (DET)}
\label{DET}

DET encrypts identical data values into identical encryptions when using the same key, i.e., $\forall x,y$:
$x =y \Leftrightarrow E(x)=E(y)$. Thus, DET allows queries with equality predicates, equi-joins, as well as \texttt{GROUP BY}, \texttt{COUNT} and \texttt{DISTINCT} queries \cite{DBLP:phd/de/Popa2014}.
DET is secure only when there is no redundancy in data. It is not robust against statistical attacks. 
Although some public key encryption schemes allow exact match queries with stronger security guarantees \cite{DBLP:conf/eurocrypt/BonehCOP04}, search takes linear time with the size of the database, while DET operates in logarithmic time \cite{DBLP:conf/crypto/BellareBO07}, thus explaining its adoption in CryptDB. 

\paragraph{Order Preserving Encryption (OPE)}
\label{OPE}

OPE is a deterministic encryption scheme that preserves plaintext order in ciphertexts. Let $x$ and $y$ be two plaintext values and $E$ an OPE scheme. If $x\leq y$, then $E(x) \leq E(y)$. This feature allows range queries, 
\texttt{MIN} and \texttt{MAX} aggregations, and ordering over ciphertexts.
In terms of security, OPE is weaker than DET because it reveals data order. 	
Yet, it can provide sufficient security for some applications, e.g., when the adversary does not possess any prior knowledge, while increasing the efficiency of query processing \cite{ozsoyoglu2003anti}.

\vspace{0.3cm}
Table \ref{tables:QueryAwareEncryptionSchemes} summarizes the features of the cryptographic schemes used in CryptDB.	
\begin{table*}
	\centering
	\caption{Features of CryptDB's encryption schemes}
	\label{tables:QueryAwareEncryptionSchemes}
	\begin{tabular}{|c|c|c|c|c|}
		\hline
		Allowed queries &                        RND& HE                                                 & DET  & OPE      \\
	
		\hline
		DISTINCT                                       & No&No                          &Yes                     & Yes        \\
		WHERE ($=$, $\neq$)                                       & No &No                          & Yes                   & Yes        \\
		Range queries                                       & No &No                          & No           & Yes         \\
		ORDER BY                                            & No&No                          & No & Yes         \\
		JOIN                                                 & No &No                          & Yes                                  & Yes        \\
		SUM, AVG                                             & No &Yes                         & No                         & No         \\
		MIN, MAX                                              & No &No                          & No & Yes                      \\
		GROUP BY                                              & No &No                          & Yes                  & Yes        \\ \hline
		Information leakage                                             & None &None                        & Duplicates &  Data order \\
		\hline         
	\end{tabular}
\end{table*}
\subsubsection{CryptDB's Architecture}
\label{Section.SecureDBs.CryptDB.Architec}

CryptDB follows three principles to solve the problem of querying encrypted databases: 1)
\textit{SQL-aware encryption} that uses cryptographic schemes within SQL queries; 2) \textit{adjustable query-based encryption} to minimize data leakage; and 3) \textit{chain cryptographic keys in user passwords} to enable data decryption only for authorized users with access privileges.

In CryptDB's core, encryption  is structured in multiple embedded levels akin to onion layers. 
Each onion layer helps process given classes of queries. 
The outermost layers are RND and HE, HE actually being Paillier's PHE scheme. RND and HE provide the highest level of security, whereas inner layers, OPE and DET, provide more functionality. 
The OPE layer is an enhancement of \cite{DBLP:journals/jdwm/boldyreva2009order}. 
Eventually, two new cryptographic schemes enable join operations.


Ciphertext access is achieved through a trusted proxy server that encrypts data, rewrites queries (by anonymizing table and attribute names and encrypting constants) and decrypts query results. 
The proxy server stores
encryption keys, the database schema and the onion layers of all attributes in the database. When a query is issued, the proxy dynamically peels off onion layers downs to a layer corresponding to the given computation. 
\hyphenation{emp-loy-ee}
For instance, consider the query \texttt{SELECT * FROM employee  WHERE  name = 'Alice'}.
First, the proxy issues a query to peel off the RND layer for attribute \texttt{name} down to the DET layer. Then, the proxy rewrites the query as 
\texttt{SELECT * FROM T1  WHERE  A2 = '0xac18f'},	
where \texttt{T1} and \texttt{A2} denote the anonymization of table \texttt{employee} and attribute \texttt{name}, respectively, and \texttt{0xac18f} = $E_{DET}($'Alice'$)$.
Similary, aggregation query \texttt{SELECT SUM(salary) FROM employee} would translate as
\texttt{SELECT SUM$_{HE}$(A3) FROM T1}, where \texttt{SUM$_{HE}$} is a user-defined function implementing Paillier's PHE and \texttt{A3} is the anonymization of attribute \texttt{salary}.

\subsection{MONOMI}
\label{Section.SecureDBs.MONOMI}

While CryptDB offers one of the first practical solutions for secure DBMSs, there are still a lot of queries that are not supported, especially OLAP-like queries.
As an illustration, CryptDB supports only 2 queries out of 22 from the TPC-H decision support benchmark \cite{tpch}. Thence, 
MONOMI builds upon CryptDB to allow the execution of analytical workloads \cite{DBLP:journals/pvldb/TuKMZ13}. 

To this aim, MONOMI adds in a designer that optimizes the physical database layout at the CSP's and a query planner that splits query execution between the CSP and the user. 
The optimal plan for executing some queries may indeed involve sending intermediate results between the user and the CSP several times to execute different parts of a query \cite{DBLP:journals/pvldb/TuKMZ13}. 
For instance, to run a \texttt{SUM / GROUP BY / HAVING} query,
MONOMI computes the \texttt{SUM} and \texttt{GROUP BY} at the CSP's through the HE and DET encryption schemes, respectively. Then, since HE does not preserve data order, the \texttt{HAVING} statement is executed at the user's after decryption. 	
This strategy helps MONOMI allow 19 out of the 22 queries of TPC-H.

\subsection{Multi-Valued Order Preserving Encryption (MV-OPE)}
\label{MV-OPE}

Lopes et al. rightly claim that ``little attention has been devoted to determine how a data warehouse hosted in a cloud should be encrypted to enable analytical queries processing'' \cite{DBLP:conf/dawak/LopesTMCC14}. Thence, they propose 
the MV-OPE scheme that allows \texttt{GROUP BY} queries over ciphertexts. Such a scheme could replace CryptDB's and MONOMI's OPE without having to compute anything at the user's.

Generally speaking, MV-OPE extends OPE by encrypting the same plaintext into different ciphertexts while preserving the order of the plaintexts \cite{DBLP:journals/ieicet/KadhemAK10}.
Thus, given two clear values $x$ and $y$ and an MV-OPE function $E$, if $x<y$ then $E(x) < E(y)$. MV-OPE can be used to compute operations such as equality, difference, inequalities, minimum, maximum and count \cite{DBLP:conf/dawak/LopesTMCC14}. MV-OPE improves robustness against statistical attacks and only leaks the order of data.   
Lopes et al.'s scheme combines MV-OPE with FHE (Section~\ref{HE}). Moreover, as CryptDB and MONOMI, it involves a secure host, e.g., a trusted proxy server. Despite using FHE, Lopes et al. experimentally show that computing queries over ciphertexts at the CSP's is significantly faster than computing them at the user's after decryption.

\subsection{Secure Trusted Hardware}
\label{Section.SecureDBs.SecureHW}

Trusted hardware devices are widely used for security, e.g., smart cards for secure authentication and secure coprocessors in automated teller machines (ATMs). Quite naturally, the idea of processing queries inside tamper-proof enclosures of trusted hardware, such as a secure coprocessor or Field Programmable Gate Array (FPGA)-based secure programmable hardware \cite{DBLP:conf/fpl/EguroV12}, came up. 
Such components are physically hosted at the CSP's. They have access to encryption keys and allow performing a limited set of queries over ciphertexts.

\subsubsection{TrustedDB}
\label{Section.SecureDBs.SecureHW.TrustedDB}

\hyphenation{pseu-do-rand-om}

TrustedDB is an SQL database processing engine that makes use of IBM 4764/5  cryptographic coprocessors \cite{DBLP:journals/ibmrd/ArnoldBCCDSHHJMW12}  to run custom queries securely \cite{DBLP:conf/sigmod/BajajS11}. Coprocessors  offer several cryptographic schemes such as the Advanced Encryption Standard (AES), the Triple Data Encryption Standard (3DES), RSA, pseu-do-random number generation and cryptographic hash functions. Yet, cryptographic coprocessors are significantly
constrained in both computation ability and memory capacity. Thus, a trade-off  must be considered between cheap query processing on untrusted main processors (at the CSP's) and expensive computation inside secure coprocessors. 

Sensitive data can only be decrypted and processed by the user or a secure coprocessor. Only non-sensitive data are stored unencrypted at the CSP's. When a query is issued, it is encrypted at the user's, rewritten as a set of subqueries and executed at the CSP's or in the secure coprocessor database engine, with respect to data sensitivity. The final result is assembled, encrypted by the secure coprocessor and sent back to the user.

\subsubsection{Cipherbase}
\label{Section.SecureDBs.SecureHW.CipherBase}

Cipherbase aims at deploying trusted hardware for secure data processing in the cloud \cite{DBLP:conf/icde/ArasuEJKKR15}. Cipherbase actually extends Microsoft SQL Server with in-server, customized FPGA-based trusted hardware. 
The FPGA is a trusted black box for computing operations over ciphertexts, which are encrypted with a non-homomorphic encryption scheme such as AES. The FPGA decrypts data internally, processes the operations and encrypts the result back. As in TrustedDB, query processing on non-sensitive data is handled by the CSP.

\subsection{Discussion}
\label{Section.SecureDBs.Disc}

CryptDB is much cited, but is quite insecure and introduces some loopholes. Its onion adjustable encryption architecture is indeed unidirectional, i.e., once an attribute is set down to a weak scheme such as DET, it never returns to a higher encryption level \cite{DBLP:conf/ccs/KerschbaumGHHKSST13}. Moreover, attributes targeted by exact match and range queries are encrypted with DET and OPE, respectively, 
and are vulnerable to statistical attacks. As a result, once an exact match or range query is issued, the system becomes vulnerable ever after. DET and OPE have even been shown to be much more insecure than
previously expected \cite{DBLP:conf/ccs/NaveedKW15}. 
Additionally, peeling down onion layers induces an overhead, especially in the case of big tables. 

Moreover, although CryptDB does support many types of queries, there are still many unsupported types of queries, e.g., predicate evaluation on more than one attribute.
MONOMI addresses this shortcoming, but retains the same security mechanisms as CryptDB. 
MONOMI also induces a heavy communication overhead between the user and the CSP, since intermediate results may be exchanged several times to execute different parts of a query \cite{DBLP:journals/pvldb/TuKMZ13}.

Despite a distributed architecture, Lopes et al's solution requires a trusted server to securely execute \texttt{GROUP BY} queries. In our database outsourcing scenario, all service providers that are external to the user's are considered untrusted. Thus, Lopes et al's trusted server would be located at the user's, inducing costs that do not fit our scenario. Additionally, this solution does not support \texttt{MIN} and \texttt{MAX} aggregation operators directly over ciphertexts. 

Finally, beside computation ability and memory capacity limitations, trusted hardware is still very expensive, which is again contrary to our scenario that aims at using cheap commodity machines in the cloud. Moreover,  
leaving unencrypted attributes jeopardizes ciphertext, because relationships between ciphertexts and plaintexts may reveal information about ciphertexts \cite{DBLP:conf/cidr/ArasuBEKKRV13}.

\section{Conclusion}
\label{section.GlobalDiscussion}

Although encryption methods enforce privacy, in some cases, the impact on performance makes them inapplicable to cloud databases. It is indeed currently impossible to develop a system  that meets both state-of-the-art cryptographic security standards and query performance requirements. In this final section, we provide a global discussion on  security, performance and storage requirements for secure databases, before concluding the paper.

\subsection{Security}
\label{section.GlobalDiscussion.Security}

The DET and OPE schemes, which are notably used in CryptDB, allow efficiently performing queries over ciphertexts.
Database optimization techniques, e.g., usual indexing methods, can also be used to enhance query performance. However, DET and OPE leak a non-negligible amount of information and are vulnerable to statistical attacks \cite{kellaris16}. 
For example, a large fraction of tuples from DET encrypted attributes can be  decrypted by statistical attacks \cite{DBLP:conf/ccs/NaveedKW15}. The vulnerability of DET is extremely detrimental to DBs with high redundancy, e.g., data warehouses. The weak security of OPE makes it inappropriate, too. It is indeed even worse than DET in terms of security \cite{DBLP:conf/cans/Furukawa14,kellaris16}.
Eventually, a recent class of generic attacks against private range query schemes invalidates much of the existing literature \cite{kellaris16}.

Thus, FHE looks like a more appropriate choice for encryption. In particular, PHE encryption can be used to sum ciphertexts, but the cost of decryption at the client's can remain high. As of today, it is indeed usually more efficient to decrypt data at the client's and then perform the aggregation, rather than processing aggregation queries over ciphertexts at the CSP's \cite{DBLP:journals/pvldb/TuKMZ13}. 
Yet, FHE is likely to become a viable alternative in the upcoming decade, with both new FHE schemes and improvements in hardware performance. However, since preserving the order of data is necessary when running queries such as sorting, grouping and range operations, the issue of designing order preserving FHE schemes will have to be addressed.

\subsection{Query Post-Processing}
\label{section.GlobalDiscussion.QueryPostProc}

\hyphenation{MON-OMI}

Tuple and table-level encryption are casually considered preferable to attribute-level encryption,  because of lower startup costs at the user's  and minimal storage costs at the CSP's \cite{hore2008managing}. However, the loss of data granularity is an important deficiency in scenarios such as OLAP. Thus, some solutions that use tuple-level encryption (Section~\ref{Section.IndexBasedMethods}) handle query processing by means of auxiliary indexes at the CSP's (e.g., bucketization-based indexing) and perform final
query processing at the user's.
Similarly, MONOMI splits the execution of queries between the user and CSP. In such solutions, it is essential to cut down the bandwidth required to transfer intermediate results and user computational resources for user side query processing \cite{DBLP:journals/pvldb/TuKMZ13}, which is quite an open issue. CPU and storage usage at the user's must indeed be minimum for maintaining the benefits of outsourcing.

\subsection{Storage Overhead}
\label{section.GlobalDiscussion.StorageOverhead}

\hyphenation{attr-ibute}

CryptDB, MONOMI and Cipherbase use attribute-level encryption, i.e., each attribute value is encrypted independently \cite{DBLP:conf/icde/ArasuEJKKR15}, 
at the cost of storage overhead. For instance, using classical AES in Cipher-Block Chaining (CBC) mode, a 32-bit integer is encrypted on 256 bits \cite{DBLP:conf/icde/ArasuEJKKR15}. Worse, Paillier's PHE scheme, which is used in CryptDB, operates over 2048-bit ciphertext \cite{DBLP:journals/pvldb/TuKMZ13}.	 
MONOMI addresses this issue by packing multiple values
from a single tuple into one PHE encryption, using Ge and Zdonik's scheme \cite{ge2007answering}. This optimization works properly for a table with many PHE-encrypted attributes, but would complicate partial updates that reset some but not all attribute values packed into a PHE tuple encryption \cite{DBLP:phd/de/Popa2014}.
Thus, although security vs. performance is necessarily a tradeoff, there is still some room for improving the storage overhead of cryptographic schemes, especially for secret sharing schemes.

\subsection{Computational Overhead}
\label{section.GlobalDiscussion.ComputationalOverhead}

\hyphenation{attr-ibutes}

Operations at the CSP's  should not involve any expensive arithmetic operations such as modular multiplication or exponentiation \cite{DBLP:reference/dbsec/Sion08}. 
However, for instance in Paillier's scheme, encrypting the sum of two clear values $x$ and $y$ requires multiplying ciphertexts $E(x)$ and $E(y)$ modulo a 2048-bit public key, i.e., $E(x+y)=E(x)\times E(y)$. Such modular multiplications are computationally expensive, especially on big tables. 

MONOMI implements a grouped homomorphic addition optimization. All to-be-aggregated attributes  
are packed in such a way that aggregation queries can be computed with a single modular multiplication. This implies that all queries must be declared ahead of time, which it is not possible for all applications, e.g., OLAP ad-hoc navigation.
Yet, performance optimization techniques, such as indexing, partitioning or view materialization, can apply onto ciphertexts. However, although they speed up some queries, they also slow down others \cite{DBLP:journals/pvldb/TuKMZ13}. As a result, it is crucial to select a cryptographic method that meets all usage constrains. Again, a tradeoff must be defined to meet the intended level of privacy while minimizing the impact on performance.
\subsection{Wrap-up}
\label{sec:WrapUp}
	
In this paper, we review the security mechanisms that can nowadays be used in the deployment of cloud databases. We particularly focus on the cryptographic schemes and the (would-be) secure systems that enable executing queries over ciphertexts without decryption. This survey highlights the potential benefits of existing solutions in a cloud computing context, but also that one must take great care about security guarantees before selecting one such solution. 
	
Moreover, cryptography cannot prevent all attacks by malicious adversaries, e.g., Distributed Denial of Service (DDoS) attacks. It is thus essential to clearly specify the objectives of cloud database deployment, to adopt security mechanisms that are adapted to these objectives. Such preliminary work shall determine the initialization of secure protocols, the choice of cryptographic schemes, the need for a trusted third party, etc. 
	
Finally, since computational performance is currently still a bottleneck, resorting to data distribution and query parallelization must be a priority. Thus, cloud frameworks such as Hadoop \cite{hadoop} 
and Spark \cite{spark} 
should be exploited in future secure cloud DBMSs.

\bibliographystyle{plain}
\bibliography{cloud_crypt}	

\end{document}